\documentclass[usenatbib,useAMS,a4paper]{mn2e}

\usepackage{txfonts}
\usepackage{epsfig}

\newcommand{\gtsimeq}{\raisebox{-0.6ex}{$\,\stackrel
        {\raisebox{-.2ex}{$\textstyle >$}}{\sim}\,$}}

\def\msun{{\rm M_{\odot}}}

\title [The transition disc frequency in M stars]
{The transition disc frequency in M stars}
\author[B. Ercolano, C. J. Clarke and T. P. Robitaille]{B. Ercolano$^1$, C. J. Clarke$^1$ and T. P. Robitaille$^2$\\
$^1$Institute of Astronomy, Madingley Rd, Cambridge, CB3 0HA, UK \\
$^2$Spitzer Postdoctoral Fellow, Harvard-Smithsonian Center for Astrophysics, 60 Garden Street, Cambridge, MA 02138, USA}

\date{Submitted: July 2103}

\pagerange{\pageref{firstpage}--\pageref{lastpage}} \pubyear{2003}

\begin{document}
\def\lta{\mathrel{\spose{\lower 3pt\hbox{$\mathchar"218$}}
     \raise 2.0pt\hbox{$\mathchar"13C$}}}
\def\gta{\mathrel{\spose{\lower 3pt\hbox{$\mathchar"218$}}
     \raise 2.0pt\hbox{$\mathchar"13E$}}}
\def\Msun{{\rm M}_\odot}
\def\msun{{\rm M}_\odot}
\def\Rsun{{\rm R}_\odot}
\def\Lsun{{\rm L}_\odot}
\def\19{GRS~1915+105}
\label{firstpage}
\maketitle

\begin{abstract}
We re-examine the recent suggestion of a high fraction of transition discs 
(i.e. those
with a cleared inner hole) in M stars, motivated by the fact that we expect
that, for M stars, even discs without inner holes should
exhibit very weak excess shortward of around $10\,\mu$m. Our analysis
of spectral energy distribution models suggest that this indeed means that M stars where a detectable
excess begins at around $6\,\mu$m may be mis-classified as transition discs
when in fact they have optically thick dust extending in to the dust
sublimation radius. Consequently, we estimate that the transition disc
fraction among M stars in the Coronet cluster is $\sim15\pm10\%$ (rather
than the  recently claimed value of $50\%$).
This revised figure would imply that the transition disc fraction is not
after all markedly higher in later type stars. We suggest that for M stars,
transition discs can only be readily identified if they have emission 
that is close to photospheric out to $> 10\,\mu$m.
\end{abstract}


\section{Introduction}
There is currently considerable interest in so-called `transition
discs' around young stars, that is systems whose spectral energy distribution (SED)
suggests that their inner regions are devoid of dust (e.g. Calvet et al 2002,
D'Alessio et al 2005, Forrest et al 2004). As the
nomenclature suggests, these objects are deemed to be in a state of
transition between disc-possessing and discless status and are of interest
inasmuch as their properties may shed light on the processes that disperse
discs around young stars.  Understanding disc dispersal  is evidently of
relevance to planet formation, whether or not, as is currently
debated,  the dispersal  interrupts  the process
or whether it is instead a signature of planet building.

  Transition discs are generally identified as objects whose SED indicates the presence of optically thick dust at
large radius (i.e. excess emission at mid infrared wavelengths or beyond)
but which do not show evidence of excess emission at shorter
infrared wavelengths. This is most readily interpreted as resulting from
an inner hole in the disc, and  transition objects have been
modeled in detail under this assumption (e.g. Rice et al 2003, Quillen et al
2004, Rice et al 2006). 
A  property of transition
discs, which has been noted since the earliest studies (e.g. Hartigan et al 1990, Kenyon and Hartmann 1995) is that they appear
to be rather rare, typically being outnumbered by about a factor ten by
objects whose discs lack  inner holes (Duvert et al 2000, Andrews and Williams
2005, Hartmann et al 2005).
A natural inference from this observation is that the timescale
for disc clearing is relatively short, so that rather few objects are
caught in the act of transition. A number of models have been motivated
by this apparent observational requirement of two timescales - i.e. an
overall disc lifetime and a substantially shorter final clearing timescale
(Clarke et al 2001, Alexander et al 2006).

  Now that a larger sample of transition objects have been identified,
thanks to the acquisition  of broad band SEDs out
to mid infrared wavelengths using  Spitzer, this conclusion appears
in general to have been upheld, with $10 \%$ still representing a rough
value for the fraction of young stars in transition
(Sicilia-Aguilar et al 2006,
Padgett et al 2006, Lada et al 2006, Hernandez et al 2007). 
This consensus was
however notably challenged by the recent study of discs in the Coronet
cluster by Sicilia-Aguilar et al 2008 (hereafter SA08) which claimed a transition disc
fraction of around $50 \%$. The authors drew attention to the
preponderance of M stars in their sample and suggested accordingly that
a lengthy transition phase may be a feature of later type stars or that some of these discs
could have been formed already as flattened or transition-like structures. 

  If substantiated, such a claim would have important implications for
disc clearing: in particular a lengthy transition phase would suggest that models that have been
developed for G and K type stars, involving rapid clearing, are only applicable
in hotter stars, a fact that could in itself lend clues about the dispersal
process. However, before exploring such theoretical implications it is necessary
to be clear that one is indeed seeing evidence for an extended transition
phase in M stars. In this Letter, we argue that the classification of
a large fraction of such stars as transition objects is a consequence
of the fact that, in the case of cooler stars, one has to go to longer
wavelengths before one can unambiguously disentangle disc emission from
that of the
stellar photosphere. Thus the SEDs seen in many
M stars that have been dubbed as transition objects (i.e. with excess
emission becoming apparent only beyond $6\,\mu$m) is actually what is to
be expected in the case of discs extending all the way in  to the dust sublimation radius: such discs obviously do produce emission at less
than $6\,\mu$m, but this is difficult to separate from that of the stellar
photosphere.

  In Section 2, we set out simple arguments why the wavelength at which one
attains a given ratio of disc to stellar emission scales, in the
case of an untruncated disc, with the reciprocal of the stellar temperature.
In Section 3 we illustrate these ideas with some specimen SEDs computed using the model database and model fitting tool
of Robitaille et al (2006, 2007). Section 4 summarises our
conclusions.

\section{Simple arguments}
We consider the case of an optically thick disc that is heated
entirely
by reprocessing of stellar radiation. We assume the star radiates as
a black body with temperature $T_*$ and the disc radiates as a sequence
of black bodies of various temperatures, $T(r)$. Radiative equilibrium
relates $T(r)$ to $T_*$ via a relation which, for the purpose of the
simple argument presented here, we paramaterise to be of the form 
\begin{equation}
T(r) \sim T_*(r/r_*)^{-q}
\end{equation}

where the value of q is dictated by the geometry of the disc surface, 
e.g. $q=0.75$ for a flat reprocessing disc and $q \rightarrow 0.5$ in the case
of a quasi-spherical distribution: observationally, the slope of the
broadband SEDs in T Tauri stars
suggests $q \sim 0.5$, which requires
the disc to be significantly flared 
(Kenyon and Hartmann 1987).  The ratio of the fluxes emitted
by the disc to that emitted by the star at wavelength $\lambda$ is thus
given by the ratio of the product of black body fluxes and emitting areas
for the star and disc. As a single temperature black body, the star's
emitting area is evidently $ \propto r_*^2$. The effective emitting area of the
disc at wavelength $\lambda$  is $\propto r_{\lambda}^2$, where $r_{\lambda}$
is the radius in the disc with 
temperature ($T_{\lambda}$)
for which  
the black
body emissivity peaks at wavelength $\sim \lambda$.   We thus have

\begin{equation}
{{L_d(\lambda)}\over{L_*(\lambda)}} = {{B_{\lambda}(T_{\lambda})}\over{B_{\lambda}(T_*)}}\biggr({{r_{\lambda}}\over{r_*}}\biggl)^2
\end{equation} 
where $B_{\lambda}(T)$ is the usual black body (Planck) emissivity. Substituting the functional form of $B_{\lambda}(T)$ and applying the
parameterisation of
$r_{\lambda}/r_*$ from (1) above, we then obtain

\begin{equation}
{{L_d(\lambda)}\over{L_*(\lambda)}} = {{\biggl(exp\bigl({{hc}\over{\lambda kT_*}}\bigr)-1
\biggr)}\over{\biggl(exp\bigl({{hc}\over{\lambda kT_{\lambda}}}\bigr)-1\biggr)}}\biggl({{T_*}\over{T_\lambda}}\biggr)^{2/q}
\end{equation}  

We have however defined $T_{\lambda}$ such that the expression containing
the exponential on the denominator is of order unity (i.e. we select the
region of the disc whose temperature means that its black body curve
is peaked at $\sim \lambda$). Therefore, since this also implies that
$T_\lambda \propto  1/\lambda$ we can write that the disc to star flux ratio
scales with $\lambda$ and $T_*$ according to\begin{equation}{{L_d(\lambda)}\over{L_*(\lambda)}} \propto  {{\biggl(exp\bigl({{hc}\over{\lambda kT_*}}\bigr)-1\biggr)}}(\lambda T_*)^{2/q}\end{equation} We thus see immediately that this 
ratio is a function of the product $\lambda T_*$ and thus that the wavelength at which a given contrast between the disc and the star is achieved is inversely proportional to the
stellar temperature. Furthermore, provided the {\it star's} spectrum is
in the Rayleigh Jeans tail at wavelength $\lambda$, the
scaling simplifies to

\begin{equation}{{L_d(\lambda)}\over{L_*(\lambda)}} \propto (\lambda T_*)^{2/q-1}\end{equation}

For $q=0.5$, the ratio thus scales with $T_*^3$. {\it Thus the disc to star ratio at fixed wavelength for a pure reprocessing disc is larger for hotter stars}. This result can be understood since, although
hotter stars are  brighter at any given  wavelength, this is more than
offset by the fact that the region of the disc contributing emission at
this wavelength is larger in the case of hotter stars. 

  The above simple argument implies that in the case of discs that extend inwards to a fixed inner temperature (e.g. untruncated discs which are optically thick as far in as the dust sublimation radius), it is
harder to detect disc signatures, at a given wavelength, in the case of
a cooler star. By the same token, if the disc is truncated so as to remove
disc emission at a given wavelength, it is harder to detect the absence
of disc emission at this wavelength. Thus it may be hard to distinguish
truncated discs and untruncated discs if the wavelength of observation is
too low. For example, if in the case of G stars one starts to detect the
presence (or absence) of excess disc emission at a few microns, then - with
given measurement errors -  it becomes possible to do the same for M stars
at a wavelength that is greater than this by the ratio of the stellar
temperatures (i.e. a factor $\sim 1.5$). 

\begin{figure}
\begin{center}
\includegraphics[width=9.0cm]{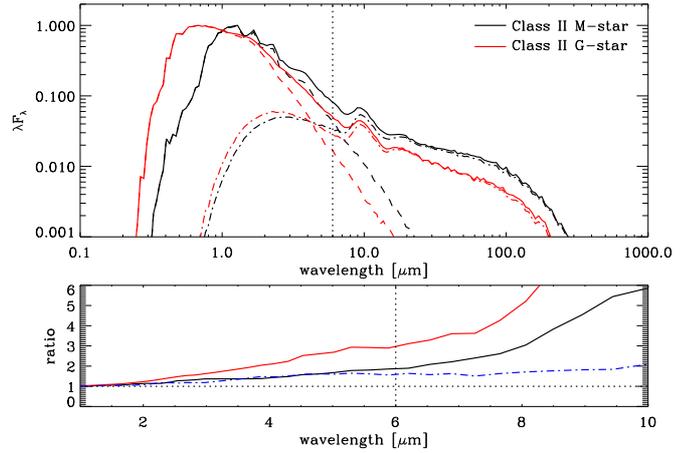}
\caption[]{{\it Top panel}: Model SEDs for a class II G star (red) and a class II M star (black) from the set of Robitaille et al. (2006). The model parameters are summarised in columns 2 and 3 of Table~1. The solid lines represent the total normalised fluxes, the dashed lines represent the normalised photospheric fluxes and the dash-dot lines represent the thermal emission. {\it Bottom panel}: Ratio of total to photospheric flux for the M star (black solid line) and G star (red solid line) models. A ratio of 1 is shown by the horizontal dotted line. The blue dash-dot line shows the total to photospheric flux ratio for the G star model divided by that of the M star model. The vertical dotted line in both panels is given to guide the eye and it marks the 6\,$\mu$m location. }
\label{f:mods}
\end{center}
\end{figure}

\begin{figure}
\begin{center}
\begin{minipage}[]{8.2cm}
\includegraphics[width=8.0cm]{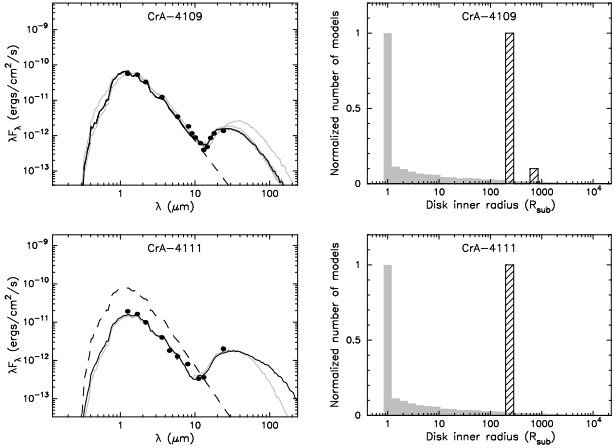}
\end{minipage}
\caption[]{{\it Left Panels}: Model fits to the digitised data points of TOs from the SA08 sample which were confirmed as bona-fide TOs by our SED fitting.  The solid and dashed black lines show the best fitting model and corresponding stellar photosphere model, while the solid grey lines show other models providing a good fit. {\it Right Panels}: The hashed histograms represent the distribution of derived values of the inner hole radius in units of the dust sublimation radius for the models which fit the SED of each object. The grey histogram shows the distribution of this quantity for all the models in the Robitaille et al. (2006) set. Both histograms are normalised so that their peak value is 1.}

\label{f:fitsTO}
\end{center}
\end{figure}

\begin{figure}
\begin{center}
\begin{minipage}[]{8.2cm}
\includegraphics[width=8.0cm]{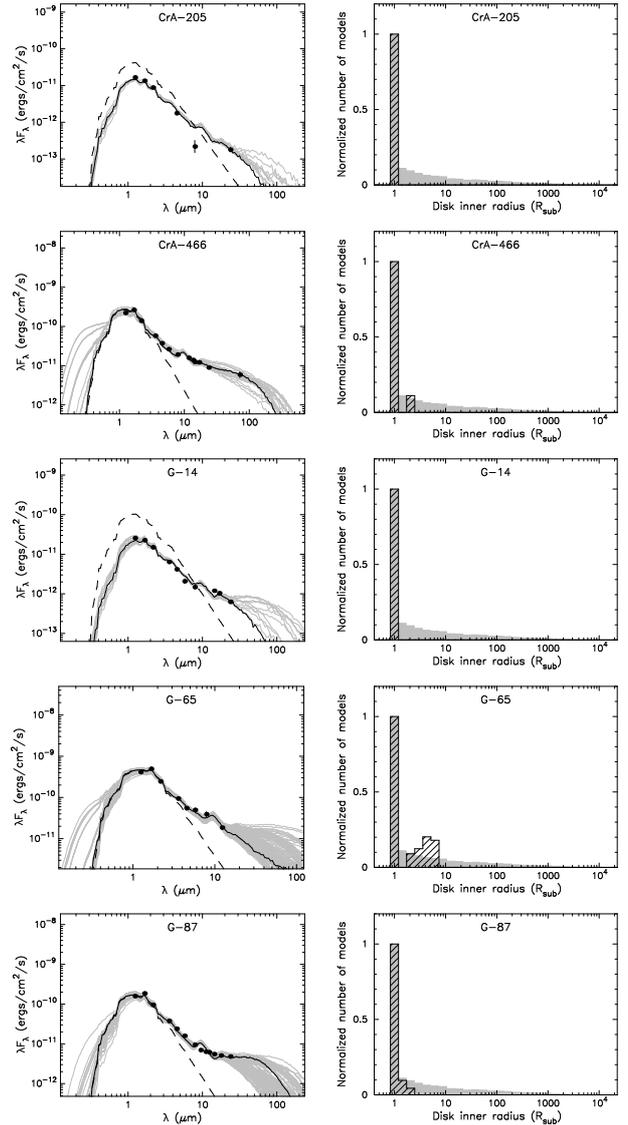}
\end{minipage}
\caption[]{{\it Left Panels}: Model fits to the digitised data points of the TOs from the SA08 sample which can be well fit by untruncated models. The lines and histograms are as in Figure 2.}
\label{f:fitsII}
\end{center}
\end{figure}

These general arguments are borne out by the SED models of Robitaille et al
(2006) shown in
 Figure~\ref{f:mods} where we have selected models for a class~II G star (red) and M star (black). The IDs of the models are 3013671 and 3009553 respectively\footnote{The exact parameter values for these models are available at http://www.astro.wisc.edu/protostars}. These models have disc accretion rates of
$\sim$5$\cdot$10$^{-10}$ $M_{\odot}/yr$; at such accretion rates the discs are in both
cases optically thick as far in as the dust sublimation radius, but in neither
case is the accretion luminosity significant. Thus these spectra correspond to those
expected for optically thick reprocessing discs.
The solid lines in the upper panel represent the total normalised flux, while the normalised photospheric flux is represented by the dashed lines. The thermal emission of the disc is also shown on the figure as the dash-dot line. The models indeed confirm that for, the cooler (T$_{\rm eff}\sim$3000\,K) M star, 
the excess above the photospheric flux at 6\,$\mu$m (marked by the vertical dotted lines) is lower than in the case of the hotter (T$_{\rm eff}\sim$5000\,K) G star. 
The ratio of total to photospheric flux is shown in the 
lower panel of Figure~\ref{f:mods}, where the black and red solid lines represent the ratios for the M- and G star models.
The blue dash-dot line shows the total-to-photospheric flux ratio for the G star model divided by that of the M star model. 
We note that at long wavelengths, this ratio
approaches the value expected in the case that both stars are fully in the 
Rayleigh Jeans limit, i.e. as given by equation (5) in the case that $q \sim 0.75$.

We thus see that this sort of SED, with small excess values shortward of $6\,\mu$m should be characteristic of reprocessing discs around M stars (of course, 
M star discs that are additionally heated by accretion may have measurable excesses at these wavelengths, thus explaining the systems that are deemed to be class~II disc systems by SA08.)  

\section{Comparison of models with data}

Here we focus on the issue of whether the objects classified by
SA08 as M star transition discs are genuinely undergoing
a phase of inner disc clearing. In Figures 2 and 3 we reproduce the SEDs of the claimed M star transition discs, which were selected
on the grounds that the emission was apparently photospheric out to
$6\,\mu$m but rose at longer wavelengths. 

 We have used the model database and fitting tool of Robitaille et al. (2006, 2007) 
in order to analyse  the SEDs of those  objects in the  Coronet cluster that were classified as 'transition objects' (TO) by SA08. The observational data were obtained by digitally sampling the SEDs shown in Figures~7 and 8 of SA08. 
The fitting tool was then used to find all models which provide a good fit to these observations,
characterising each fit by a $\chi^2$ value. In the case of five of the sources, a spectral type is available (SA08). For these sources, we reject models which differ by more than one spectral subclass. For example, for CrA-466, classified as M2, we reject all models later than M3 and earlier than M1.
The range of parameters of the models providing a good fit (which we define as $\chi^2-\chi_{\rm best}^2 < 2\,n_{\rm data}$, where $\chi_{\rm best}$ is the $\chi^2$ of the best fitting model and $n_{\rm data}$ is the number of data points) are in each case given in Table 1;
in Figures 2 and 3 the solid black line represents the SED of the best-fit model,
the dashed black line represents the spectrum of the central source, and the solid black lines show other models providing a good fit according to the above criterion.

From our SED fitting, we found that only two of the seven claimed TO's cannot be fit by models with discs extending all the way to the dust sublimation radius, and hence show evidence of a cleared inner hole. Figure~\ref{f:fitsTO} 
(left panels) shows the model fits ($\chi^2-\chi_{\rm best}^2 < 2\,n_{\rm data}$) to the digitized data points of the bona-fide TOs confirmed by SED fitting. The right-hand panels (hashed histogram) 
depict the distribution of derived
values of the inner hole radius in units of the dust sublimation radius,
for the subset of models that provide a good fit to the
SED of each object; this is compared (grey histogram) with the distribution of this
quantity for all the 
2$\cdot 10^5$ models in the grid. Figure~\ref{f:fitsII} shows SED fits and disc inner radius distributions for the objects that had been 
previously classified as TOs by SA08, but turned out to be well-fit by untruncated 
disc models. We also note that our fit for CrA-205 shown in Figure~\ref{f:fitsII} does not go through the 8\,$\mu$m IRAC point. 
SA08, however, report an error of 0.38 magnitudes  on the 8\,$\mu$m IRAC point and an uncertain measurement for the 24\,$\mu$m MIPS point: furthermore the IRS spectra published for this object in the 8\,$\mu$m region has a very low signal-to-noise.

On this basis, we estimate that the true
fraction of M stars in the Coronet that are demonstrably in transition (i.e. have 
discs where there is good evidence for truncation at a radius exceeding
the dust sublimation radius) is closer to $\sim2/13=15\%$ (with an uncertainty of $\sqrt{2}/13=10\%$ due to small number statistics).

Finally, we note that the models that fit the observations all require
that the dust distribution is at most modestly flared and that the dust is
apparently settled relative to the gas distribution. This is shown in the last two columns of Table 1, which list the range of values of the scaleheight of the disc at 100\,AU divided by the scaleheight the disc would have if it was in hydrostatic equilibrium with $T \propto r^{-0.5}$ (which would imply $h \gtsimeq r^{5/4}$, D'Alessio et al. 1997, Chiang \& Goldreich 1997).
Such a tendency towards flatter discs (i.e. with a more settled  dust 
component) in lower mass stars has been noted by a number of previous authors
(Pascucci et al 2003, Apai et al 2004, Allers et al 2006).  

\begin{table*}
\caption[]{Parameters of the models presented in Figures 2 and 3.}
\label{t:t1}
\begin{tabular}{lcccccccccc}
\hline
& \multicolumn{2}{c}{$M_{\rm disc}$ ($M_{\odot}$)} & \multicolumn{2}{c}{$\dot{M}_{\rm disc}$ ($M_{\odot}$/yr)}  & \multicolumn{2}{c}{$R_{\rm min}$ ($R_{\rm sub}$)} &  \multicolumn{2}{c}{$h$ at 100\,AU (AU)} &  \multicolumn{2}{c}{$h/h_{\rm HSEQ}$ at 100\,AU} \\
Source name  & min & max & min & max & min & max & min & max & min & max\\
\hline
CrA-205    &     2.7$\times10^{-8}$ &     3.8$\times10^{-4}$ &     8.5$\times10^{-14}$ &     2.7$\times10^{-10}$ &        1.0 &        1.2 &        1.6 &        8.1 &        0.2 &        0.5 \\
CrA-466    &     3.3$\times10^{-4}$ &     9.5$\times10^{-3}$ &     4.5$\times10^{-10}$ &     2.3$\times10^{-8}$ &        1.0 &        2.2 &        2.0 &        8.0 &        0.3 &        0.7 \\
CrA-4109   &     6.8$\times10^{-7}$ &     2.1$\times10^{-5}$ &     5.5$\times10^{-13}$ &     2.0$\times10^{-11}$ &      232.5 &      596.3 &        2.3 &        4.3 &        0.3 &        0.5 \\
CrA-4111   &     2.2$\times10^{-6}$ &     1.7$\times10^{-4}$ &     2.1$\times10^{-11}$ &     4.0$\times10^{-10}$ &      200.3 &      220.9 &        2.6 &        3.1 &        0.3 &        0.3 \\
G-14       &     3.5$\times10^{-7}$ &     7.8$\times10^{-4}$ &     2.2$\times10^{-13}$ &     1.8$\times10^{-10}$ &        1.0 &        1.1 &        3.9 &        7.5 &        0.4 &        0.6 \\
G-65       &     1.6$\times10^{-6}$ &     1.5$\times10^{-2}$ &     3.9$\times10^{-13}$ &     1.7$\times10^{-8}$ &        1.0 &        5.5 &        1.5 &        7.5 &        0.3 &        0.7 \\
G-87       &     1.6$\times10^{-6}$ &     4.0$\times10^{-3}$ &     3.8$\times10^{-13}$ &     5.2$\times10^{-9}$ &        1.0 &        1.9 &        1.6 &        8.7 &        0.3 &        0.6 \\
\hline
\end{tabular}

\end{table*}     

\section{Conclusions}

We have shown that in the case of pure reprocessing discs, the contribution
from the disc is small in M stars, shortward of $6\,\mu$m, even 
if the disc extends all the way in to the dust sublimation radius.
We can understand this based on 
the simple arguments presented in Section 2: although the 
photospheric output is lower in cooler stars, this is more than offset 
by the fact that the area of disc emitting at a given wavelength is lower, 
and in consequence the contrast between the thermal and stellar contribution 
(disc to star ratio) at given wavelength is lower than in the case of a hotter star. 
This means that reprocessing 
discs around M stars will tend to have the sorts of SEDs typified by Cr-466 (see Figure 3), being close to
photospheric at $< 6\,\mu$m, but evidencing excess emission at longer wavelength.
Such discs are therefore not necessarily in transition. Spectral models
(Figure 2) suggest that one only starts to get unambiguous evidence of holes
in M star discs in cases where the SED is close
to photospheric out to $> 10\,\mu$m. Thus an object like CrA-4109 
is likely to contain an inner
hole. Based on these arguments we suggest that the transition disc frequency
for M stars in the Coronet has been over-estimated and that, as a preliminary
estimate, $15\pm10\%$  might be a closer figure.  

The high-temperature analogue of this effect was discussed by Whitney et al. (2004), who found 
that the infrared excess from the disc appears enhanced compared to
cooler stars, resulting in a misidentification of hotter sources as
younger (Class I) sources. As shown by their models, a range of stellar temperatures all result
in the same shape for their thermal spectrum, and, as we discussed
earlier, the difference in the apparent excess is due to the contrast
between the stellar and thermal emission. Whitney et al. (2004)
describe the behavior as a separation in the wavelengths of the peak
emission between the stellar and thermal emission, which is much
larger in hotter stars, resulting in the thermal emission being much
more evident in the 1-10\,$\mu$m range. 

Another effect that should perhaps be mentioned here is that M star luminosities are 
generally lower than G star luminosities, hence their dust sublimation radius is on 
average smaller. In some cases one might expect that the dust sublimation radius may 
reside inside the magnetic truncation radius, as predicted by magnetic accretion models. 
In such cases one might also detect an an infrared hole, even though the object is not a transition disc. 
This effects does indeed hint at the possibility that M star clusters may exhibit a larger apparent TO ratio. 
Our analysis of the Coronet cluster, however, implies a TO fraction of $15\pm10\%$, only marginally larger 
than those observed for G stars. 

 Finally, we note that our discussion appears to imply that inner holes are 
harder to detect in M stars than in G stars. This conclusion however
needs to be qualified, since this refers to the detectability of 
truncation according to the presence or absence of excess {\it at a 
given wavelength}.  We noted in Section 2 that the wavelength at which
the disc to star ratio equals a given value is inversely proportional to
temperature, and therefore one can test for truncation in lower mass stars
by going to longer wavelengths. In brief, if one's 
measurement errors allow one to detect a given ratio of disc to
photospheric emission, then it allows one to detect holes which deplete
dust that is hotter  
than a fixed fraction of the stellar temperature.
Given the form of the disc termperature profile (equation 1) this minimum
detectable hole then corresponds to a fixed multiple of the stellar radius.
Therefore, if one has observations extending to long enough wavelength,
one can detect physically smaller holes in (physically smaller) cooler stars.
We suggest, on inspection of the models shown in Section 3, that in M stars
one can safely label a disc as `transitional' only if the upturn in flux
occurs at wavelengths longer than $10\,\mu$m, 
but emphasise that more detailed modeling is
required to make this figure precise.

\section{Acknowledgments}
We thank the referee Barbara Whitney for helpful discussion and suggestions which
aided to the clarity of our work. We also thank Aurora Sicilia-Aguilar for a
thorough analysis of our work and for helpful suggestions. TPR acknowledges
support from NASA through the Spitzer Space Telescope Fellowship Program (TPR).




\begin{thebibliography}{}
\bibitem[]{} Alexander, R., Clarke, C., Pringle, 2006. MNRAS 369,229
\bibitem[]{} Allers, K., Kessler-Silacci, J., Cieza, L., Jaffe, D., 2006. ApJ 644,364
\bibitem[]{} Apai, D., Pascucci, I., Sterzik, M., van der Bliek, N., Bouwman, J., Dullemond, C., Henning, T., 2004. A \& A L426,53
\bibitem[]{}Calvet, N., D'Alessio, P., Hartmann, L., Wilner, D., Walsh, A., Sitko, M., 2002.ApJ 568,1008
\bibitem[Chiang 
\& Goldreich(1997)]{1997ApJ...490..368C} Chiang, E.~I., \& Goldreich, P.\ 1997, ApJ, 490, 368 
\bibitem[]{}Clarke, C., Gendrin. A., Sotomayor, M, 2001. MNRAS 328,485
\bibitem[]{}D'Alessio, P., et al, 2005. Ap J 621,461
\bibitem[D'Alessio et al.(1997)]{1997ApJ...474..397D} D'Alessio, P., 
Calvet, N., \& Hartmann, L.\ 1997, ApJ, 474, 397 
\bibitem[]{}Duvert, G., Guilloteau. D., Menard, F., Simon, M., Dutrey, A., 2000. A \& A 355,165
\bibitem[]{}Forrest, W. et al , 2004. ApJS 154,443
\bibitem[]{}Hartigan, P., Hartmann, L., Kenyon, S., Strom, S., Skrutskie, M., 1990. Ap J 354,L25
\bibitem[]{}Hartmann, L., Megeath, T., Allen, L., Luhmann, K., Calvet, N., D'Alessio, P., Franco-Hernandez, R., Fazio, G., 2005. ApJ 629,881
\bibitem[]{}Hernandez, J, et al 2007. ApJ 671,1784
\bibitem[]{}Kenyon, S., Hartmann, L., 1987. ApJ 323,714
\bibitem[]{}Kenyon, S., Hartmann, L., 1995. ApJS 101,117
\bibitem[]{}Lada, C. et al, 2006, AJ 131,1574
\bibitem[]{}Padgett, D., et al , 2006. Ap J 553,383
\bibitem[]{} Pascucci, I., Apai, D., Henning, T., Dullemond, C., 2003. ApJ L590,111
\bibitem[]{}Quillen, A., Blackman, E., Frank, A., Varniere, P., 2004. ApJ 612,L137
\bibitem[]{}Rice, W., Lodato, G., Pringle, J., Armitage, P., Bonnell, I., 2004. MNRAS 355,543
\bibitem[]{}Rice, W., Wood, K., Armitage, P., Whitney, B., Bjorkmann, J., 2003. MNRAS 342,79
\bibitem[Robitaille et al.(2006)]{2006ApJS..167..256R} Robitaille, T.~P., Whitney, B.~A., Indebetouw, R., Wood, K., \& Denzmore, P.\ 2006, ApJS, 167, 256 
\bibitem[Robitaille et al.(2007)]{2007ApJS..169..328R} Robitaille, T.~P., Whitney, B.~A., Indebetouw, R., \& Wood, K.\ 2007, ApJS, 169, 328 
\bibitem[]{}Sicilia-Aguilar, A., Hartmann, L., Calvet, N., Megeath, T., Muzerolle, J., Allen, L., D'Alessio, P., Merin, B., Stauufer, J., Young, E., Lada, C., 2006. ApJ 638,897
\bibitem[]{}Sicilia-Aguilar, A., Henning, T., Juhasz, A., Bouwman, J., Garmire, G., Garmire, A., 2008 (SA08)
\bibitem[Whitney et al.(2004)]{2004ApJ...617.1177W} Whitney, B.~A., 
Indebetouw, R., Bjorkman, J.~E., \& Wood, K.\ 2004, ApJ, 617, 1177 



\end{thebibliography}
\end{document}